\documentclass[runningheads]{llncs}

\usepackage[T1]{fontenc}
\def\doi#1{\href{https://doi.org/\detokenize{#1}}{\url{https://doi.org/\detokenize{#1}}}}
\usepackage{graphicx}
\usepackage{url}            
\usepackage{booktabs}       
\usepackage{amsfonts}       
\usepackage{nicefrac}       
\usepackage{microtype}      
\usepackage{xcolor}         
\usepackage{listings}
\usepackage{algorithm} 
\usepackage{algpseudocode} 

 \usepackage[sort,comma,numbers]{natbib}

\begin{document}
\title{BrainActivity1: A Framework of EEG Data Collection and Machine Learning Analysis for College Students}
%
%
\titlerunning{BrainActivity1}
%
\author{Zheng Zhou\inst{1}\orcidID{0000-0001-9313-2106} \and
Guangyao Dou\inst{1}\orcidID{0000-0001-8011-9658} \and
Xiaodong Qu\inst{1,2}\orcidID{0000-0001-7610-6475}}

\authorrunning{Z. Zhou, G. Dou, X. Qu}
%
\institute{Brandeis University, Waltham MA 02453, USA\\ \email{\{zhengzhou, guangyaodou\}@brandeis.edu}
\and Swarthmore College, Swarthmore PA 19081, USA\\
\email{xqu1@swarthmore.edu}}

%
%
\maketitle              

\begin{abstract}
Using Machine Learning and Deep Learning to predict cognitive tasks from electroencephalography (EEG) signals has been a fast-developing area in Brain-Computer Interfaces (BCI). However, during the COVID-19 pandemic, data collection and analysis could be more challenging than before. This paper explored machine learning algorithms that can run efficiently on personal computers for BCI classification tasks. Also, we investigated a way to conduct such BCI experiments remotely via Zoom. The results showed that Random Forest and RBF SVM performed well for EEG classification tasks. The remote experiment during the pandemic yielded several challenges, and we discussed the possible solutions; nevertheless, we developed a protocol that grants non-experts who are interested a guideline for such data collection.

\end{abstract}
\section{Introduction}

Previous research in Computer Science, Neuroscience, and Medical fields has implemented EEG-based Brain-Computer Interfaces (BCI) in several ways, \cite{lotte2018review,craik2019deep,qu2020identifying,appriou2020modern,lotte2010regularizing, lotte2014tutorial,qu2020multi,qu2020using,jamil2021cognitive,darvishi2021neurophysiological}, such as diagnosis of Alzheimer's, emotion recognition, mental workload, motor imagery tasks \cite{devlaminck2010circular,lotte2015signal,lotte2015towards,bashivan2014spectrotemporal,bashivan2015learning,bashivan2016mental}. Machine learning, deep learning, and transfer learning algorithms have demonstrated the great potential in such biomarker data analysis \cite{lotte2007review,zhang2020survey,xu2020multi,gu2020multi,roy2019deep,miller2019library,kaya2018large,zhao2020sea,lotte2018bci,bird2018study, zhao2021mt,qian2021two,bhat2019ultra,basaklar2021wearable,dongare2021categorization}.

 \begin{figure}[!t]
\centering
  \includegraphics[width=0.69\textwidth]{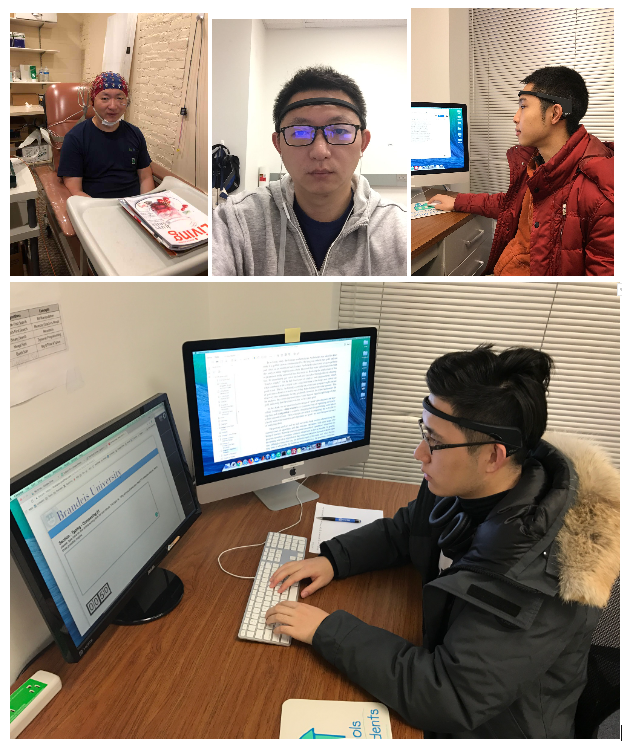}
  \caption{College Students using EEG Devices}
  \label{fig:four_in_one}
\end{figure}

However, EEG datasets' size is still relatively small compared with peers in Computer Vision and Natural Language Processing \cite{lotte2018review,qu2018eeg,craik2019deep}. 
 
Our \textbf{research questions} are: 
1. Can we develop a larger dataset with data from a larger audience? for example, college students?
2. Can we design easy-to-use BCI experiments to collect EEG data with consumer-grade devices for college students?
3. Can we develop a step-by-step guide for such a data collection and machine learning analysis process?

\section{Methods}
 
As \cite{ienca2018brain,qu2018personalized,portillo2021mind,cannard2021validating} mentioned, several affordable (less than three hundred dollars) non-invasive consumer-grade EEG headsets are commercially available. As shown in Figure \ref{fig:four_in_one}, we have pilot-tested several clinical and non-clinical EEG devices with college students. The Top left cell in Figure \ref{fig:four_in_one} showed an example of the wearing of a clinical device while others demonstrated the consumer-grade devices. Muse Headset were used as an example for demonstration, followed by more details  in the 6-page full-length page.
 
\begin{figure}[!t]
\centering
  \includegraphics[width=0.9999\textwidth]{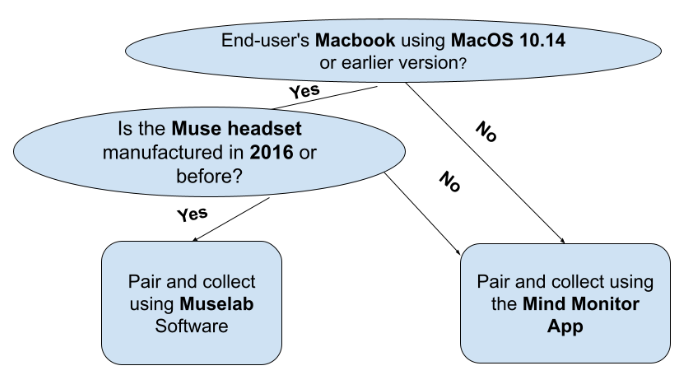}
  \caption{Data Collection App Selection}
  \label{fig:data_collection_app}
\end{figure}

\textbf{As shown in Figure} \ref{fig:data_collection_app}, first, we installed the data collection software or application. The following method section was formulated based on the example of using Muse Headset. We have used this device since 2016. First, the end user's OS version must match the version of the EEG recording application. If the end-user had a newer Macbook or the latest version of Muse headset, they ought to use the Mind Monitor Application to record EEG signals. Otherwise, the end-user may still use the Muselab application for EEG recording. 
  
\begin{figure}[!b]
\centering
  \includegraphics[width=0.9999\textwidth]{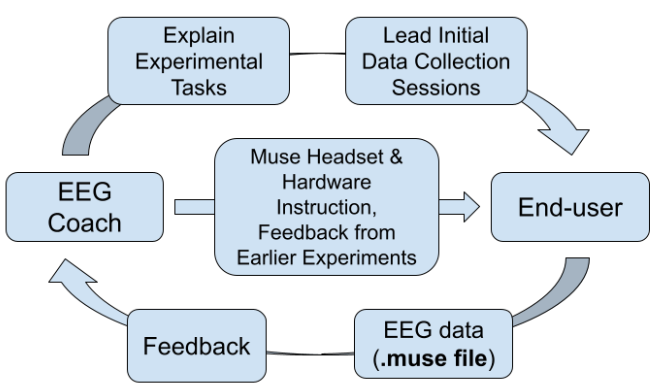}
  \caption{Data Collection Flow Chart, User and Coach}
  \label{fig:data_collection_user_coach}
\end{figure}

 \begin{figure}[!t]
\centering
  \includegraphics[width=0.69\textwidth]{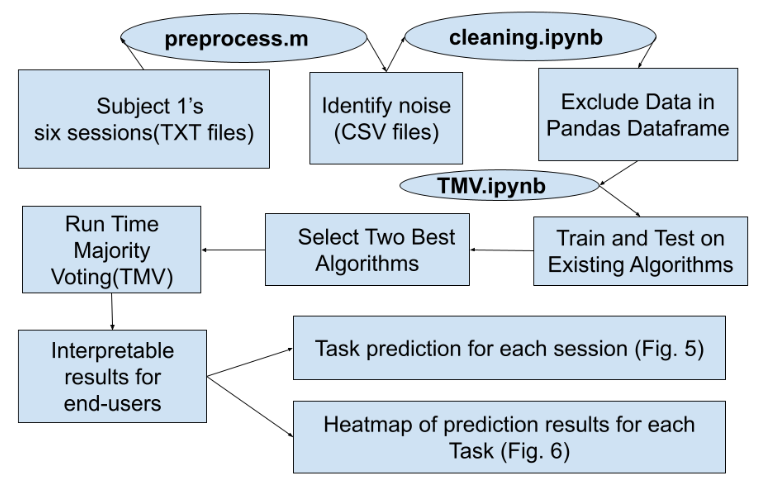}
  \caption{Data Analysis Flow Chart}
  \label{fig:EEG_Data_analysis_flow_chart}
\end{figure}

\begin{figure}[!b]
\centering
  \includegraphics[width=0.78\textwidth]{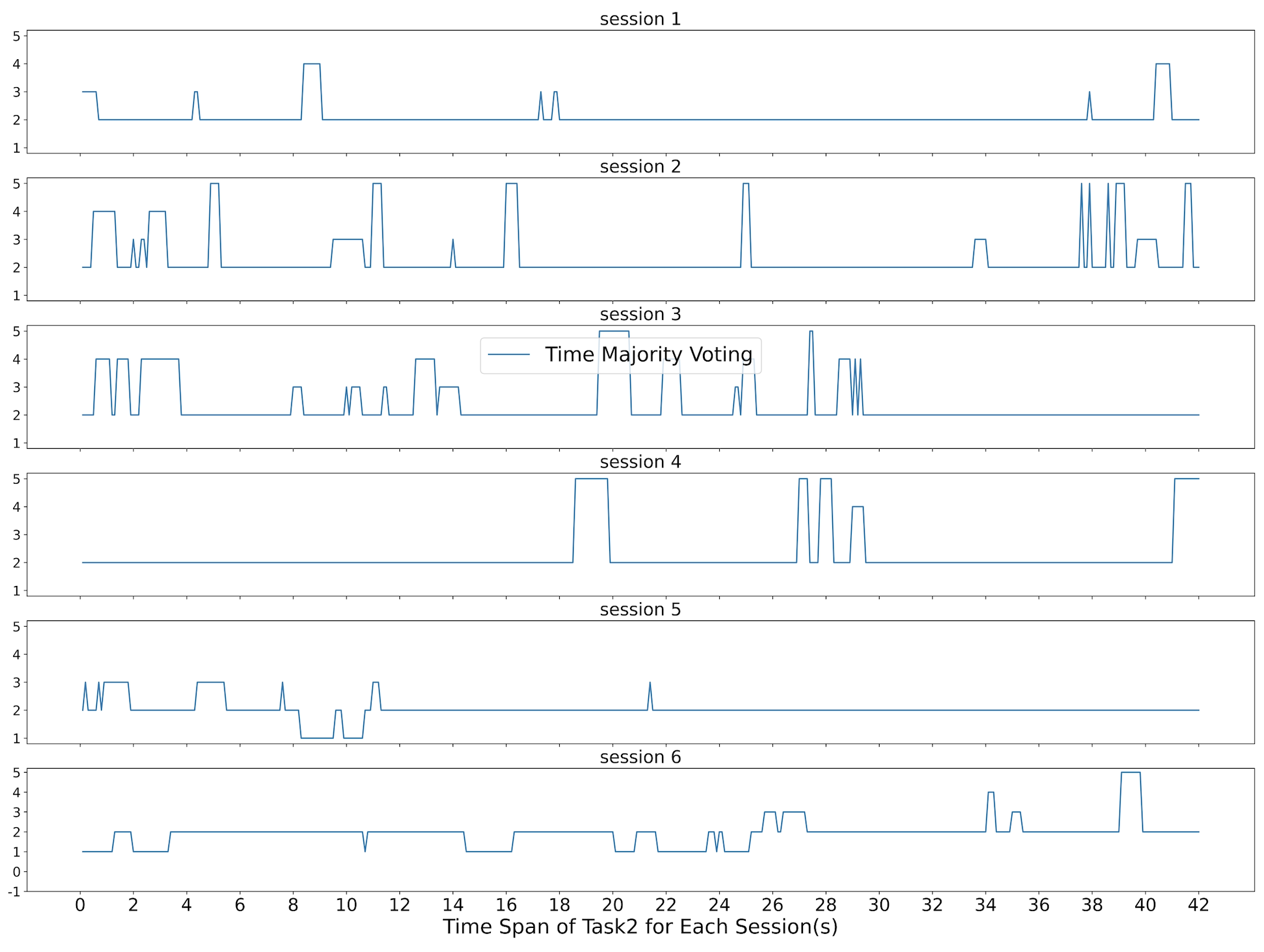}
  \caption{Subject 1 Task 2 Prediction for all Six Sessions}
  \label{fig:subject_1_task_1_prediction}
\end{figure}

Two individuals, the EEG coach, and the end-user, were usually required to complete the data acquisition of such non-invasive EEG signals via Muse headset and Muse Recording software, As shown in Figure \ref{fig:data_collection_user_coach}. First, the EEG Coach introduced the Muse Headset and Muse recording application to the end-user. Then the EEG coach explained the details of a specific experiment. Next, the EEG coach and the end-user started the experiment to collect the data. The data was saved as a MUSE file. The end-user then summarized the experimental feedback to the EEG coach. Together with the research team, the EEG coach generated visual feedback to the End-user. Such feedback may contribute to the development of future experiments. 

If the end-user was interested, they could learn the data analysis themselves, which took two hours on average for students who major in computer science.
  
Most EEG recording applications came with a toolset to convert the recording files to TXT or CSV files. Afterward, we could pick the subset of data we planned to use for further analysis; we recommended starting with the absolute value of the EEG signals.
 
 We implemented several machine learning algorithms commonly used in the field \cite{breiman1996bagging,breiman2001random,breiman2017classification,chevalier2020statistical} from the scikit-learn \cite{scikit-learn}. For example, Linear Classifiers, Nearest Neighbors, Decision Tree, and Ensemble Methods.  
 
 As shown in the Figure \ref{fig:EEG_Data_analysis_flow_chart}, once we had six TXT files for all six sessions of data, we first executed a Matlab program - preprocess.m - to identify the noises and turn all the TXT files into separate CSV files. Then, we ran the Clean.ipynb to exclude these data in the Pandas Dataframe for further analysis. Next, we executed the TMV.ipynb to train and cross-validate existing machine learning algorithms such as the Random Forest, SVM, and KNN. Then we selected the top two best-performing algorithms and used these two algorithms to perform Time Majority Voting. In our case, these two algorithms were Random Forest and RBF SVM. Lastly, the TMV.ipynb would generate a visualization of task predictions for end-users.  Figure \ref{fig:subject_1_task_1_prediction} and a heatmap for all six sessions (Fig \ref{fig:subject_1_TMV_heatmap}). These two figures are examples of task one’s prediction results for all six sessions of subject 1 in the TCR Experiment.

\section{Results}
 
As shown in Figure \ref{fig:subject_1_task_1_prediction}, sessions designed task were clearly recognized. All six sessions of subject 1's task 2 showed consistent patterns. Such visual feedback was provided to the EEG coach and end-user who collected this set of data. From the experiment notes, we ran into a signal issue behind the right ear of the end-user throughout the session, then we recorded this case and potential solutions to improve future experiments.
 
\begin{figure}[!t]
\centering
  \includegraphics[width=0.95\textwidth]{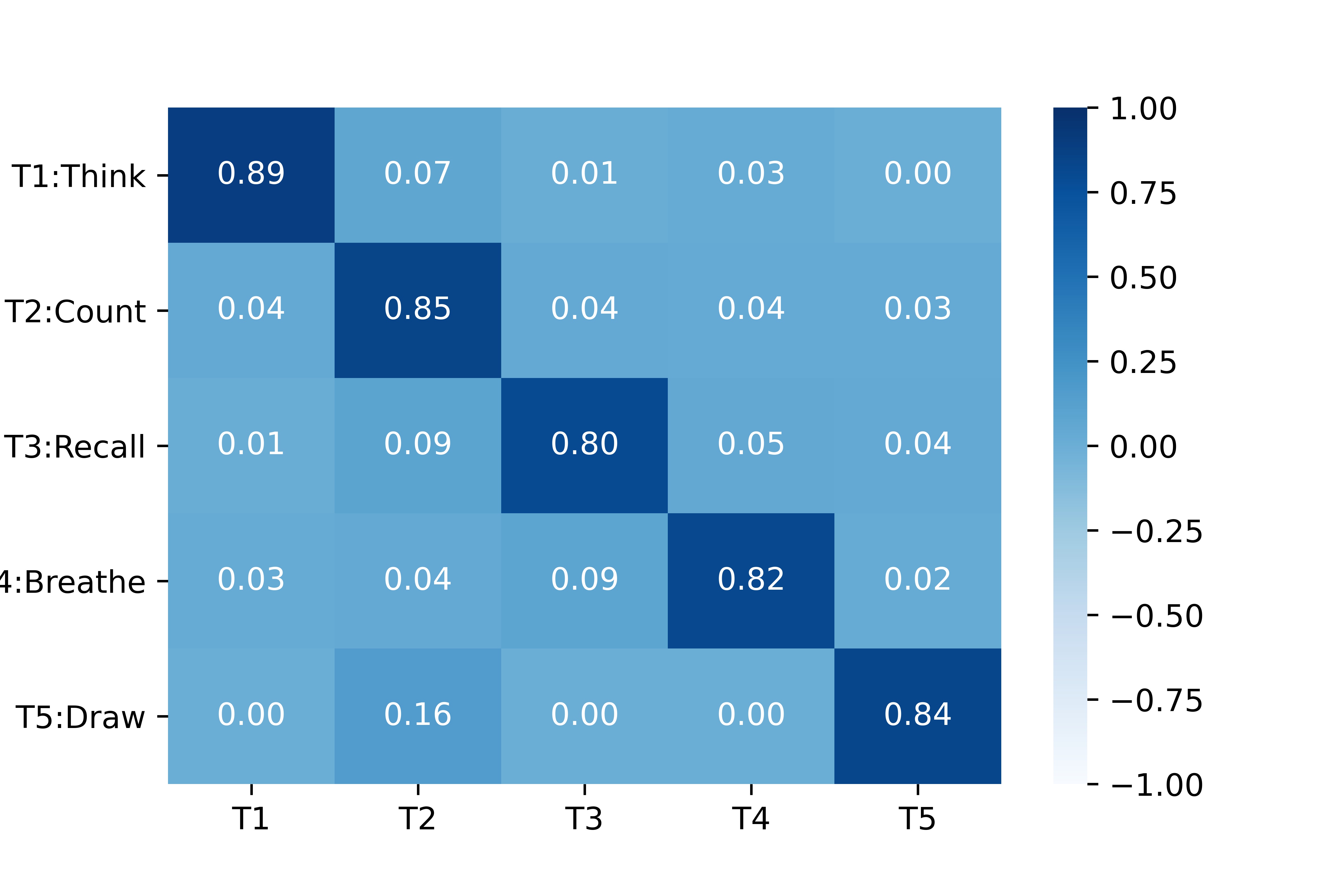}
  \caption{Subject 1's TMV Heatmap for all Six Sessions}
  \label{fig:subject_1_TMV_heatmap}
\end{figure}

As shown in Figure \ref{fig:subject_1_TMV_heatmap}, the X-Axis is the designed tasks, the Y-Axis is the predicted tasks. The diagonal means the designed tasks matched the prediction. Such visual feedback also helped the research team and the end-users better understand what task pairs were easy to be confused with each other.

\section{Discussion}
This paper proposed an approach for non-export independent researchers to collect EEG-based BCI data with affordable non-clinical devices. When performing test trials with Muse headsets, we provided a general guideline, as shown in figure \ref{fig:data_collection_app} and \ref{fig:data_collection_user_coach}, which showed promising result in many EEG data collections performed by naive EEG end-users. This general guideline demonstrated a decision tree for non-expert researchers to acquire data collection hardware and software. We also presented our data collection flow, which formed a closed loop between the researchers and the experimental subjects. In addition, we elaborated our data-cleaning analysis procedures. 

Significant progress has been made in user-training for EEG-based BCI studies, while the framework proposed in this paper serves as a stepping stone for further improved training programs in future research. However, some limitations were identified along the course of our project. Our project spanned from pre-pandemic to post-pandemic time. We found that in-person data collection trials were significantly more efficient than trials that took place virtually during the global pandemic. We ought to explore more strategies and updated methods that could grant us the efficiency when data collection has to be completed in a virtual environment.

Even though as much detail and trial and error experience we managed to include in our guideline, there are chances that individual cases develop distinct issues. Our future work includes
research/EEG coach-based student study/work community, in which they can learn and discuss their experience with non-clinical devices collecting EEG-based data and possibly establish solutions to various issues after the encounter.

\section{Conclusion}
This paper investigated the data collection for EEG-based BCI to develop larger datasets. We explored the possibility of collecting EEG data from college students with affordable devices. The results demonstrated that the proposed framework could simplify the process and contribute to developing a larger EEG dataset.

%
%
%
%
\makeatletter
\renewcommand\@cite[2]{%
  Ref.~#1\ifthenelse{\boolean{@tempswa}}
    {, \nolinebreak[3] #2}{}
}
\renewcommand\@biblabel[1]{#1.}
\makeatother
 \bibliographystyle{splncs04}
 \bibliography{data}
\end{document}